\documentclass[3p,twocolumn]{elsarticle}

\usepackage{latexsym}
\usepackage{amssymb}
\usepackage{graphicx}  
\usepackage{epsfig}    
\usepackage{dcolumn}   
\usepackage{bm}        

\usepackage{color}

\begin{document}

\title{High-quality ion beams from nanometric double-layer targets and their application to hadron-therapy}

\author[pks]{M.~Grech\corref{cor1}} \ead{mickael.grech@gmail.com}
\author[pks,jena]{S.~Skupin}
\author[cea]{R.~Nuter}
\author[cea]{L.~Gremillet}
\author[cea]{E.~Lefebvre}

\cortext[cor1]{Corresponding author}
\address[pks]{Max Planck Institute for the Physics of Complex Systems, D-01187 Dresden, Germany}
\address[jena]{Institute of Condensed Matter Theory and Optics, Friedrich Schiller University, D-07743 Jena, Germany}
\address[cea]{CEA, DAM, DIF, F-91297 Arpajon, France}

\begin{abstract}
The production of ion beams from the interaction of a circularly polarized laser pulse with a nanometric double-layer target is discussed in the regime where all electrons are expelled from the target by laser radiation pressure. Quasi-monochromatic, well-collimated ion beams are observed in two-dimensional particle-in-cell simulations.  The ion beam properties are derived from a simple analytical model, and the possibility to control those properties by using a laser-pulse with sharp-rising edge is discussed. Potential application to hadron-therapy is finally considered.
\end{abstract}

\maketitle

\section{Introduction}\label{sec_intro}

Generating high-energy ion beams by irradiating a solid target with an ultra-intense laser pulse has many potential applications from medicine to controlled nuclear fusion~\cite{ion_applications}. While different mechanisms of ion acceleration have been demonstrated, ion acceleration at currently available laser intensities mainly follows from the plasma expansion driven by the hot, laser-created, electrons~\cite{tnsa}. Experimentally, protons with energy up to 70~MeV have already been produced~\cite{snavely_robson}. However, controlling their energy distribution remains a crucial problem and alternative acceleration mechanisms are considered to overcome this issue~\cite{dce_theory,dce_exp,rpa,teravetisyan}.

In this paper, we revisit ion acceleration from a double-layer target and propose a new laser-based acceleration scheme that allows an excellent control of the ion beam properties. This mechanism relies on the complete expulsion of electrons from an ultra-thin double-layer target by laser radiation pressure. The target is almost instantaneously transformed into a capacitor-like structure, where the ions of the first layer and the electron cloud play the role of the anode and cathode, respectively. A strong, quasi-homogeneous, electrostatic field  ($\sim 100\,{\rm TV/m}$) is built up between the two layers. The ions of the second layer, with a larger charge-over-mass ratio, are then accelerated in a way similar to ions in a conventional linear accelerator. Later on, once electrons escape from the ion layers, light ions can gain further energy in the electrostatic field of the bare first ion layer.

Two-dimensional (2D) particle-in-cell (PIC) simulations and an analytical model show that high-energy ($> 100$~MeV) ion beams can be generated over a few laser periods only. These beams are characterized by small and adjustable energy ($\lesssim 10\,\%$) and angular ($\lesssim 5^{\circ}$) dispersions, which makes them suitable for applications such as hadron-therapy.


\section{Ion beams from a double-layer target}

\subsection{Basic picture of ion acceleration}

Double-layer targets for laser-based ion acceleration have recently attracted a lot of attention as they allow to control the energy dispersion of the accelerated ions~\cite{dce_theory,dce_exp}. The first target layer is usually made of heavy ions with mass $m_h \gg m_p$ ($m_p=1836$ is the proton mass) and charge $Z_h$. The second target layer (or a small dot placed at the rear-side of the first layer) consists of ions with a larger charge-over-mass ratio ($Z_l/m_l > Z_h/m_h$). In what follows, we denote by $n_h$, $d_h$ and $w_h$ the atomic density, thickness and transverse size of the first layer, respectively. Similarly, $n_l$, $d_l$ and $w_l$ denote the corresponding parameters for the second target layer or structured dot (if $w_l < w_h$). For the sake of simplicity, all quantities in this paper are expressed in normalized units: masses, charges, densities and distances are normalized to the electron mass $m_e$, electron charge $e$, critical density $n_c = \epsilon_0\,m_e\,\omega_L^2/e^2$ ($\epsilon_0$ is the permittivity of vacuum and $\omega_L$ is the laser frequency) and inverse laser wave number $k_L^{-1}=c/\omega_L$ ($c$ is the light velocity), respectively. In addition, the laser field amplitude $a_L$, as well as all other electric fields, are given in units of the Compton field $E_C = m_e\,\omega_L\,c/e$.

When irradiated by an intense laser pulse ($a_L \gg 1$), the target (or a part of the target) is quickly ionized. Electrons are expelled from the target, and a strong electrostatic field is built-up. Light ions of the second layer are accelerated in this field. The condition for generation of quasi-monochromatic, well-collimated, light ion beams is simply that the accelerating electrostatic field remains quasi-homogeneous over the second target layer during the acceleration process~\cite{dce_theory,dce_exp}. In this paper, we revisit ion acceleration from a double-layer target in the regime where the laser radiation pressure effectively expels all electrons from the target.

\subsection{2D PIC simulations}\label{sec_2dpic}

\begin{figure*}\centering
\includegraphics[width=14cm]{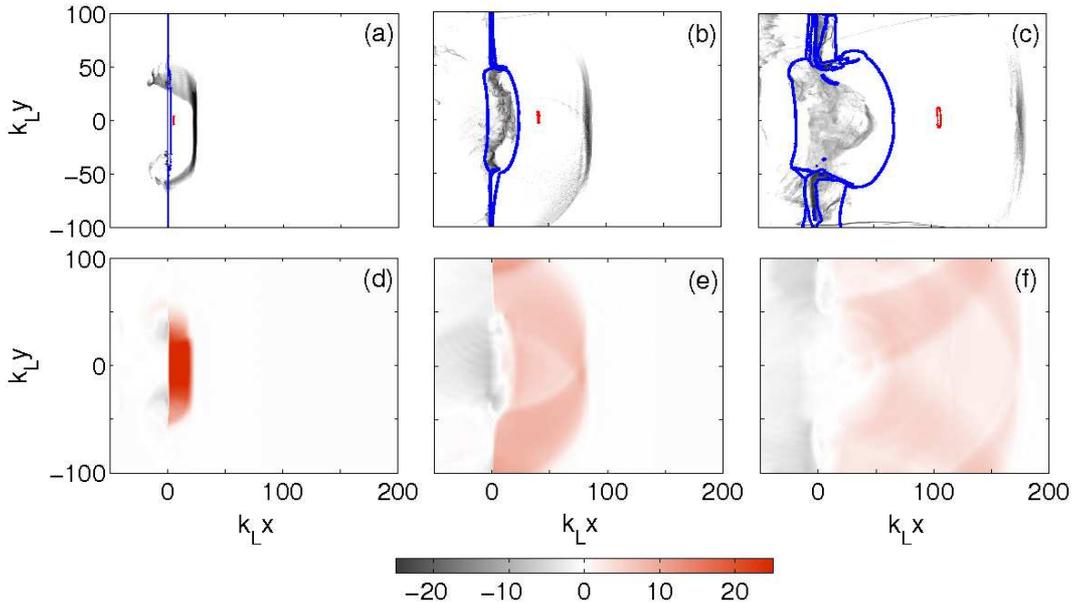}
\caption{(a,b,c) Snapshots of the electron density (gray scale), carbon contour plot (blue) and proton contour plot (red), and (d,e,f) corresponding electrostatic field distribution at three different times after the beginning of the interaction: (a,d) 4 optical cycles; (b,e) 14 optical cycles and (c,f) 30 optical cycles. The colormap refers to panels (d-f). In this simulation, $w_L=62.8$ (10 laser wavelengths) and $n_l=6$.}
\label{fig1}
\end{figure*}

\begin{figure}[h!]\centering
\includegraphics[width=7cm]{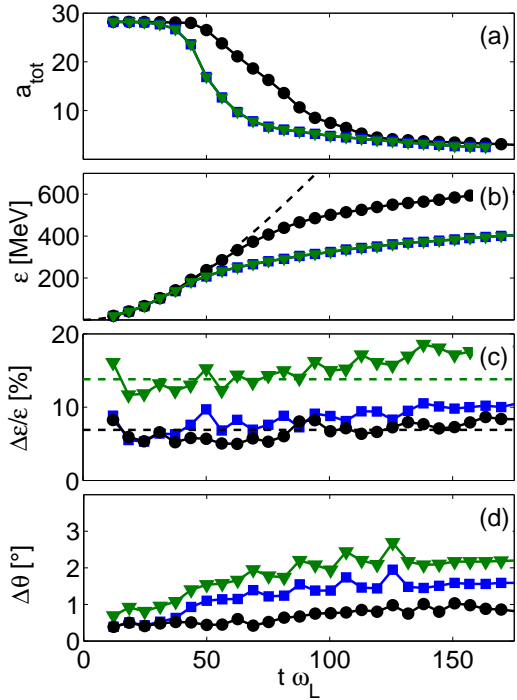}
\caption{Temporal evolution of: (a) the accelerating electrostatic field $a_{e.s.}$ at the position of the proton bunch; (b)~the proton energy $\epsilon$; (c)~the energy dispersion $\Delta\epsilon/\epsilon$; (d)~the angular aperture $\Delta\theta$. For $n_l=6$, $w_L=10\,\lambda_L$ (blue, square), $n_l=12$, $w_L=10\,\lambda_L$ (green, triangle) and $n_l=6$, $w_L=20\,\lambda_L$ (black, circle). Dashed curves account for analytical predictions of Sec.~\ref{sec_model}. Note that in panels~(a) and~(b) the blue and green curves are superimposed.}
\label{fig2}
\end{figure}

Two-dimensional simulations are performed using the PIC code {\textsc CALDER}~\cite{calder}. A laser pulse with field amplitude $a_L=100$ is focused at normal incidence on an ultra-thin ($d_h,d_l \ll 1$) double-layer target. A circularly polarized laser pulse with flat-top temporal intensity profile is used in order to make electron sweeping by the laser radiation pressure more efficient~\cite{macchi_PRL_05,kulagin}. The pulse duration is $\tau_p \sim 62.8$ (10 optical cycles) only. Along the transverse $y-$direction, the laser intensity follows a 6-th order super-Gaussian profile. Two values for the laser focal-spot full-width at half-maximum have been considered: $w_L=62.8$ and $125.6$ (10 and 20 laser wavelengths, respectively). The first target layer is made of fully ionized carbon ($Z_h=6, m_h=12\,m_p$) with atomic density $n_h = 58$ and thickness $d_h=0.08$. Its transverse size is $w_h \sim 200$ ($\sim$ 30 laser wavelengths), which is also the transverse size of the simulation domain. A likewise fully ionized hydrogen dot ($Z_l=1$, $m_l=m_p$) with transverse size $w_l=6.28$ (one laser wavelength) and thickness $d_l=0.16$ is placed at the rear-side of the first layer. Two atomic densities, $n_l=6$ and $n_l=12$, have been considered for the hydrogen dot. The simulations results are summarized in Figs.~\ref{fig1} and~\ref{fig2}. 

Four laser-periods after the beginning of the interaction, all electrons in a region with diameter $w_{\perp} \sim 70 \sim w_L$ irradiated by the laser pulse have been removed from the target (Fig.~\ref{fig1}a). In contrast to previous works~\cite{dce_theory,dce_exp} where electrons are strongly heated and generate an accelerating sheath at the target rear-side, electrons here form a compact bunch that propagates with the front of the laser pulse. The target then exhibits a capacitor-like structure and a quasi-homogeneous electrostatic field is built up between the bare heavy ion layer and the electron bunch (Fig.~\ref{fig1}d). Light ions are accelerated in this electrostatic field and can reach high energies in a few optical cycles only. In Fig.~\ref{fig2}b, the light ion energy $\epsilon$ is shown as a function of time. Four optical cycles after the beginning of the interaction, protons have already reached $\epsilon \sim 70\,{\rm MeV}$ (blue line, square). 

Due to the quasi-homogeneity of the accelerating field, both relative energy dispersion $\Delta\epsilon/\epsilon \sim 10\,\%$ and aperture angle $\Delta\theta \lesssim 5^{\circ}$ remain small (Figs.~\ref{fig2}b and~\ref{fig2}c). Moreover, both of them are, at this stage of the acceleration process, sensitive to the charge density in the proton dot: the electrostatic repulsion between the protons themselves is therefore the main source of energy and angular dispersion. 
 
This early stage of proton acceleration is maintained as long as the distance between the electron cloud and the bare heavy ion layer remains small compared to their transverse size $w_{\perp}$. At the end of this stage (at $t \sim 50$ for the smaller laser focal spot and $t \sim 80$ for the larger one), protons with energies of the order of 200~MeV and 480~MeV, respectively, have been produced. Later on, the electrostatic field seen by the protons strongly decreases due to geometrical effects (Figs.~\ref{fig1}b,c,e,f and~\ref{fig2}a) and the acceleration process slows down (Fig.~\ref{fig2}b). 

Figure~\ref{fig2}b also shows that, while the proton energy in the early stage of ion acceleration does not depend on the transverse width $w_L$ of the laser focal spot, increasing $w_L$ allows to enhance the final proton energy by delaying multi-dimensional effects. Furthermore, this later stage is associated with a slight increase of both the energy and angular dispersion of the proton bunch (Figs.~\ref{fig2}c and~\ref{fig2}d). This is the sign that both dispersions result from electrostatic repulsion in the proton dot itself, which evolves on a time scale different from that of the acceleration process. Nevertheless, for the lower hydrogen density ($n_l=6$), energy and angular dispersion seem to saturate at the end of the simulation to values not exceeding $10\,\%$ and $3^{\circ}$, respectively. Finally, we point out that while $\Delta\epsilon/\epsilon$ is only slightly sensitive to the transverse size of the laser pulse, $\Delta\theta$ is decreased for the larger focal spot. This prompts us to suggest that the transverse inhomogeneity of the accelerating field also plays a role in the proton angular dispersion.

\section{Analytical model}\label{sec_model}

On the contrary to previous studies~\cite{dce_theory,dce_exp}, the acceleration mechanism discussed here relies on the complete and directed expulsion of all electrons from the double-layer target. This occurs if the laser radiation pressure $\sim a_L^2$ is sufficient to overcome both the electron thermal pressure, and the maximal electrostatic pressure $a_{tot}^2/2$ resulting from complete electron-ion separation. The electrostatic field $a_{tot}=a_h+a_l$, where $a_h=Z_h\,n_h\,d_h$ and $a_l=Z_l\,n_l\,d_l$, is the electrostatic field between the bare ion layers and the electron cloud. Using a circularly polarized laser pulse allows to maintain the electron temperature to a rather low level~\cite{macchi_PRL_05}, so that one obtains the simple condition for complete electron removal $a_L^2 > a_{tot}^2/2$. This condition defines a threshold for the laser field amplitude $a_L> a_{tot}$ beyond which electrons are removed from the target and propagate as a compact bunch at the front of the laser pulse~\cite{kulagin}.

\subsection{Linear plasma acceleration}

Previous numerical simulations show that early light ion acceleration (typically for $t < w_L$) is one-dimensional (1D). It occurs in a monotonically increasing accelerating field that preserves the order of the different ion sheets that make up the light ion layer. Light ions therefore see a constant accelerating field whose amplitude depends only on their initial position. The light ion mean energy $\epsilon$ and energy dispersion $\Delta\epsilon$ (in units of $m_e\,c^2$) can thus be computed exactly~\cite{grech_NJP_09}:
\begin{eqnarray}
\label{eq1} \epsilon_{linPA}       &=& m_l\,\left[ \sqrt{1+t^2/t_r^2} - 1\right]\,, \\ \nonumber
\label{eq2} \Delta\epsilon_{linPA} &=& m_l\,\left[ \sqrt{1+\big(1+a_l/a_h\big)^2\,t^2/t_r^2} \right.\\
                           & & -\left. \sqrt{1+t^2/t_r^2}\right]\,,
\end{eqnarray}
where the index linPA accounts for linear plasma acceleration (with obvious analogy to conventional acceleration techniques), and $t_r= m_l/(Z_l\,a_h)$ denotes the time over which light ions gain relativistic energies. For $t \ll t_r$, they behave classically and the mean energy and relative energy dispersion are $\epsilon^{(c)} \sim m_l\,t^2/(2\,t_r^2)$ and $(\Delta\epsilon/\epsilon)^{(c)} \sim 2\,a_l/a_h$, respectively. If light ions experience the field $\sim a_h$ over a time $t \gg t_r$, they gain ultra-relativistic energies $\epsilon^{(ur)}\sim m_l\,t/t_r$, and their relative energy dispersion simply reads $(\Delta\epsilon/\epsilon)^{(ur)} \sim a_l/a_h$. 

These analytical estimates are compared to the 2D simulation results of Sec.~\ref{sec_2dpic}. In the first stage of the acceleration process, light ions see a constant accelerating field $a_{tot} \sim 28$ (Fig.~\ref{fig2}a). The resulting temporal evolution of the ion energy $\epsilon$ as predicted by Eq.~(\ref{eq1}) is in excellent agreement with the numerical results during the first acceleration stage (Figs.~\ref{fig2}b). A good agreement is also found between the predicted value of $\Delta\epsilon / \epsilon$ and numerical results (Fig.~\ref{fig2}c). In particular, $\Delta\epsilon/\epsilon \propto a_l/a_h$ is kept small by ensuring that the accelerating field $a_h$ remains large compared to $a_l$, which determines the ratio of the areal charges of both ion layers, and limits the brightness of the resulting ion source.

\subsection{Long time behavior and DCE}

The 1D behavior of light ion acceleration breaks down once the electron cloud is separated from the heavy ion layers by a distance of the order of its transverse size $w_{\perp} \sim w_L$. Because electrons leave the target at a velocity close to $c$, this defines a characteristic time $t_{linPA} \sim w_{\perp} \sim w_L$ over which the capacitor-like acceleration is maintained. On longer times, electrons eventually escape from the ion layers, and their contribution to the later stage of the ion acceleration process can be neglected. Light ions thus can gain further energy in the electrostatic field due to the bare heavy ion layer in a way similar to the so-called directed Coulomb explosion (DCE)~\cite{dce_theory}. Assuming $Z_h/m_h \ll 1$, the energy gained during this later stage can be derived from the conservation of the total light ion energy:
\begin{eqnarray}
\label{eq3} \epsilon_{DCE}       \sim Z_l\,a_h\,w_{\perp}/4 - \epsilon_{linPA}/2\,.
\end{eqnarray}
The first term in the right-hand-side of Eq.~(\ref{eq3}) is similar to estimates obtained in Ref.~\cite{dce_theory}. The second term accounts for the fact that, in contrast to what occurs in ``classical'' DCE, part of the light ions potential energy has already been transformed into kinetic energy in the first acceleration stage. 

As suggested by our numerical simulations, energy dispersion during this later stage mainly follows from the electrostatic repulsion between the accelerated light ions. A first attempt to estimate this effect is given in Ref.~\cite{grech_NJP_09}, however, the method used there is not correct. Here, we propose to estimate the energy dispersion by considering the ion bunch at the end of its acceleration stage as an homogeneously charged sphere with radius $R \sim w_l/2$, expanding due to its own charge $Q_l$ (in units of $e\,n_c/k_L^3$). The final energy dispersion can then be derived as a function of the final ion energy $\epsilon$:
\begin{eqnarray}
\label{eq4} \Delta\epsilon_{DCE} \sim 4\,\sqrt{\epsilon\,\epsilon_p}\,,
\end{eqnarray}
where $\epsilon_p=Z_l\,Q_l/(2\pi\,w_l)$ is the initial potential energy of a light ion on the outer shell of the sphere.

\subsection{Scaling laws}\label{scaling}

A question arises from previous calculations: which of the two accelerating stages provides the main contribution to the light ion acceleration? By comparing Eqs.~(\ref{eq1}) and~(\ref{eq3}), one obtains that ions with the final energy $\epsilon \lesssim m_l/8$ gain most of their energy during the later, DCE-like, acceleration stage. For these ions, the final energy is $\epsilon \sim Z_l\,a_h\,w_{\perp}/4$ and energy dispersion $\Delta\epsilon/\epsilon \sim 4\,\sqrt{\epsilon_p/\epsilon}$ follows mainly from Coulomb repulsion. Both quantities can be controlled by adjusting the target parameters. Moreover, since $w_{\perp} \sim w_L$, the condition for electron removal, $a_L>a_h$, defines a threshold for the laser power and we obtain that the light ion energy scales as $\epsilon \propto Z_l\,\sqrt{P_L}/4$.

In contrast to that, more energetic ions (in particular relativistic ions) quickly separate from the heavy ion layer so that they gain most of their energy during the early, linPA, stage. Using $t_{linPA} \sim w_{\perp} \sim w_L$ and $a_L > a_h$, Eq.~(\ref{eq1}) can be rewritten as a function of the laser power. One then obtains a threshold power, $P_L > m_l^2/Z_l^2$, above which relativistic ions with energy $\epsilon \propto Z_l\,\sqrt{P_L}$ are expected.

\section{Application to hadron-therapy}

Hadron-therapy is certainly one of the most appealing applications of laser-created ion beams. It typically requires ions with energy of the order of 200~MeV/nucleon. For such energies, $m_l/8 < \epsilon < m_l$, both stages of the acceleration process are important (see Sec.~\ref{scaling}). The final energy and energy dispersion are thus obtained from Eqs.~(\ref{eq1}) and~(\ref{eq3}) and Eqs.~(\ref{eq2}) and~(\ref{eq4}), respectively. 
We now give the characteristic parameters of a target designed for generation of high-quality proton or carbon ion beams. For given ion energy (let us say 200~MeV per nucleon) and charge-over-mass ratio, the quantity $a_h\,w_{\perp}$ is fixed. Considering $w_{\perp} \sim 62.8$ (10 laser wavelengths), we obtain that accelerating field amplitudes $a_h \sim 16$ and $a_h \sim 32$ are required to create the proton and carbon ion beams, respectively. For proton acceleration, a nanometric ($\sim 5$~nm-thick considering a $1~{\rm \mu m}$-laser wavelength), fully-ionized, diamond-like carbon foil at solid density can be used as a first layer to generate such an accelerating field. For carbon ion beam generation, a gold target is here considered. Assuming partial ionization of the gold atoms ($Z_h \sim 50$), the required target thickness is $\sim 1$~nm. The second ion layer transverse width $w_l \sim 12.6$ (two laser wavelengths) is chosen $\ll w_{\perp}$ in order to minimize the transverse inhomogeneity of the accelerating field. Considering $\Delta\epsilon/\epsilon \sim 2\%$, we finally obtain from Eqs.~(\ref{eq2}) and~(\ref{eq4}) that Coulomb repulsion between the accelerated ions limits the accelerated charge $Q_l$ to $\sim 0.4$~pC for protons and $\sim 0.9$~pC for carbon ions per shot. We would like to stress that this limitation is generic to short, non-neutral ion bunches. 

In order to ensure that the laser radiation pressure is sufficient to remove all electrons from the target, laser intensities above $3.5\times 10^{20}\,{\rm W/cm^2}$ and $1.4\times 10^{21}\,{\rm W/cm^2}$ (correspondingly laser powers exceeding 360~TW and 1.5~PW, and laser energies of a few tens of J) are required depending on whether one considers proton or carbon ion acceleration. To accelerate a sufficient number of ions, repetition rates in the kHz-domain are necessary.

\section{Conclusion}\label{sec5}

In summary, the generation of quasi-monochromatic, well-collimated, light ion beams from a nanometric double-layer target is discussed in the regime where the laser radiation pressure is sufficient to remove all electrons from the target. The resulting ion acceleration, which is shown to occur in two steps, is described using 2D PIC simulations and an analytical model. Both, the light ion source energy and energy dispersion are shown to depend mainly on the target design. On the contrary to previous findings, we show that the light ion source energy dispersion depends mainly on the electrostatic repulsion in the accelerated bunch itself, which limits the source brightness.

\section*{Acknowledgments}

Fruitful discussions with V.~T.~Tikhonchuk, A.~Macchi and T.~Cowan are acknowledged.


\end{document}